\documentclass[prd,nofootinbib,floats,superscriptaddress,eqsecnum,tightenlines,11pt]{revtex4}
\pdfoutput=1
\usepackage{hyperref}
\usepackage{graphicx}
\usepackage{amsmath,amssymb,amsfonts,amsthm,latexsym,stmaryrd}
\usepackage{marginnote}
\usepackage{color}

\newcommand{\nb}{\nonumber}
\def\beq{\begin{equation}}
\def\eeq{\end{equation}}
\newcommand{\bea}{\begin{eqnarray}}
\newcommand{\eea}{\end{eqnarray}}
\def\bfig{\begin{figure}}
\def\efig{\end{figure}}

\begin{document}

\title{Cosmological evolution in DHOST theories}

\author{M. Crisostomi}
\affiliation{Institut de physique th\'eorique, Universit\'e Paris Saclay, CEA, CNRS, 91191 Gif-sur-Yvette, France}
\affiliation{AIM, CEA, CNRS, Universit\'e Paris-Saclay, Universit\'e Paris Diderot, Sorbonne Paris Cit\'e, F-91191 Gif-sur-Yvette, France}
\affiliation{Laboratoire de Physique Th\'eorique, CNRS, Universit\'e Paris-Sud, Universit\'e Paris-Saclay, 91405 Orsay, France}
\author{K. Koyama}
\affiliation{Institute of Cosmology and Gravitation, University of Portsmouth, Portsmouth, PO1 3FX, UK}
\author{D.  Langlois}
\affiliation{Laboratoire Astroparticule et Cosmologie, CNRS, Universit\'e Paris Diderot, 75013, France}
\author{K. Noui}
\affiliation{Institut Denis Poisson, CNRS, Universit\'e de Tours, 37200 Tours, France}
\affiliation{Laboratoire Astroparticule et Cosmologie, CNRS, Universit\'e Paris Diderot, 75013, France}
\author{D.A.~Steer}
\affiliation{Laboratoire Astroparticule et Cosmologie, CNRS, Universit\'e Paris Diderot, 75013, France} 

\date{\today}

\begin{abstract}
\noindent In the context of 
Degenerate Higher-Order Scalar-Tensor (DHOST) theories, we study cosmological solutions and their stability properties.  In particular, we explicitly illustrate  the crucial role of degeneracy by showing how the higher order homogeneous equations in the physical frame (where matter is minimally coupled) can be recast in a system of equations that do not involve higher order derivatives.
We study the fixed points of the dynamics, finding the conditions for having a de Sitter attractor at late times. 
Then we consider the coupling to matter field (described 
for convenience
by a k-essence Lagrangian) and find the conditions to avoid gradient and ghost instabilities at linear
order in cosmological perturbations, extending previous work.  Finally, we apply these results to a simple 
subclass 
of DHOST theories, showing  that  de Sitter attractor conditions, no ghost and no gradient instabilities conditions (both in the self-accelerating era and in the matter dominated era) can be compatible. 
\end{abstract} 

\maketitle

\section{Introduction}
Since the discovery of the accelerated expansion of our universe, many models of dark energy or modified gravity have been proposed to account for this unexpected observation (see \cite{Clifton:2011jh, Joyce:2014kja, Koyama:2015vza, Heisenberg:2018vsk} for reviews). Among these models, scalar-tensor theories of gravity have played a prominent role as they simply add a scalar degree of freedom to the usual tensor modes of general relativity. In order to have a general understanding of the impact of scalar-tensor theories on cosmology and on astrophysics, it is  convenient to resort to a unified approach that can describe  as many models as possible with the same formalism. The most general framework  that has been developed so far is that of Degenerate Higher-Order Scalar-Tensor (DHOST) theories~\cite{Langlois:2015cwa,BenAchour:2016fzp} 
(see also \cite{Crisostomi:2016czh,Achour:2016rkg}), 
which extends the family of Horndeski~\cite{Horndeski:1974wa} (also known as generalized galileons~\cite{Nicolis:2008in,Deffayet:2011gz,Kobayashi:2011nu}) and Beyond Horndeski (or GLPV) theories~\cite{Gleyzes:2014dya, Gleyzes:2014qga}.

DHOST theories allow for the presence of second-order derivatives of the scalar field $\phi$, i.e. of $\nabla_\mu\!\nabla_\nu\phi$ in the Lagrangian, as in Horndeski theories. However, in contrast to the latter, which are restricted to Lagrangians leading to second-order Euler-Lagrange equations (for both the metric and the scalar field), DHOST theories allow for higher-order Euler-Lagrange equations but are required to contain only one scalar degree of freedom  in order to avoid Ostrogradski instabilities, associated with an extra degree of freedom that often appears in systems with higher order time derivatives.
The possibility of having higher order Euler-Lagrange equations without an extra degree of freedom was illustrated by disformal transformations of the Einstein-Hilbert Lagrangians \cite{Zumalacarregui:2013pma} and by Beyond Horndeski (or GLPV) Lagrangians~\cite{Gleyzes:2014dya, Gleyzes:2014qga}. 

It was later realized that the crucial property shared by these models is the degeneracy of their Lagrangian, which guarantees the absence of a potentially disastrous extra degree of freedom~\cite{Langlois:2015cwa}. 
The absence of an extra degree of freedom was confirmed, for Beyond Horndeski theories, by their relation to Horndeski theories via field redefinition~\cite{Gleyzes:2014qga,Crisostomi:2016tcp} as well as a Hamiltonian analysis for a particular quadratic case~\cite{Deffayet:2015qwa}, and for quadratic DHOST theories, by a general Hamiltonian analysis \cite{Langlois:2015skt}\footnote{For a systematic analysis of higher derivative theories and the conditions required to avoid Ostrogradski instabilities, see \cite{Motohashi:2016ftl, Klein:2016aiq} for classical mechanics and \cite{Crisostomi:2017aim} for field theories.}.
In the case of Beyond Horndeski theories, let us stress that  the degeneracy restricts the possible combinations of quadratic and cubic terms \cite{Langlois:2015cwa, Crisostomi:2016tcp}, meaning that a special tuning between the free functions of Beyond Horndeski theories is required \cite{BenAchour:2016fzp}.

\smallskip

The purpose of this work is to explore the cosmology of DHOST theories, first at the level of background evolution and then for linear perturbations. Some preliminary investigations were conducted in \cite{Crisostomi:2017pjs} for a special class of DHOST theories (see  \cite{DeFelice:2010pv, DeFelice:2010nf} for a similar analysis in Horndeski theories and \cite{Kase:2018iwp} in Beyond Horndeski theories). 
In the present work, we illustrate explicitly the crucial role of degeneracy by showing how the higher order homogeneous equations in the physical frame (where matter is minimally coupled) can be recast in a system of equations that do not involve higher order derivatives. This  second system corresponds to another frame, which we call Horndeski frame, where matter is non-minimally coupled. The transition from one frame to 
the other
 is given by a disformal transformation of the metric, similarly to the conformal transformation that relates the physical (or Jordan) and Einstein frames  in traditional scalar-tensor theories. 

On using the homogeneous cosmological evolution equations, we  study the existence of self-accelerating solutions, and then  compute the conditions for these fixed points to be attractors. We also construct scaling solutions which interpolate between the matter and the accelerating eras. We then study linear cosmological perturbations about these different solutions. 
Using the effective theory of dark energy perturbations applied to DHOST theory \cite{Langlois:2017mxy} (based on and extending the previous works \cite{Gubitosi:2012hu,Gleyzes:2013ooa,Gleyzes:2014rba}), we obtain the Lagrangian governing the linear degrees of freedom, thus identifying
 the conditions that guarantee the linear stability of the system. Finally, we apply this general analysis to a simple illustrative example of a DHOST theory. 

The outline of the paper is the following. In the next section, we introduce DHOST theories, focusing on the phenomenologically viable subclass (quadratic theories with stable linear perturbations), which contains Horndeski and Beyond Horndeski theories, and we describe  the homogeneous cosmological dynamics in terms of the scale factor of the physical frame or that
of the Horndeski frame.   Then, at the end of  section \ref{section_DHOST}, we identify self-accelerating solutions and investigate whether or not they are attractors  with respect to  the homogeneous cosmological evolution. Linear cosmological perturbations are studied in section \ref{section_perts}: we consider first the pure scalar-tensor action and then include a matter action consisting of a k-essence type scalar field. Section \ref{section_application} is devoted to the application of our results  to a simple and illustrative family of models. We give a conclusion in the final section.

\section{DHOST theories and homogeneous cosmological equations}
\label{section_DHOST}

We start with the general quadratic DHOST action \cite{Langlois:2015cwa,Crisostomi:2016czh,Achour:2016rkg}
\bea
\label{dhost_action}
S[g_{\mu\nu},\phi] =  \int d^4x \sqrt{-g} \left[ 
F_0(\phi,X) + F_1(\phi,X) \Box \phi + F_2(\phi,X) R + \sum_{I=1}^5 A_I(\phi,X) L_I^{(2)} 
\right] \,,
\eea
where $g_{\mu\nu}$ is the metric to which matter is minimally coupled, i.e. $g_{\mu\nu}$ corresponds to the physical frame (or Jordan frame) metric. Furthermore, $X\equiv g^{\mu\nu} \partial_\mu \phi \, \partial_\nu \phi$, and the last term in (\ref{dhost_action}) contains all five possible Lagrangians $ L_A^{(2)} $, quadratic in second derivatives of the field. They are given by
\bea
&&L_1^{(2)} \equiv  \phi^{\mu \nu}\phi_{\mu \nu}, \qquad L_2^{(2)} \equiv  (\phi_{\nu}{}^{\nu})^2, \qquad L_3^{(2)} \equiv  \phi_{\nu}{}^{\nu} \phi^\rho \phi_{\rho \sigma} \phi^{\sigma} \,,
\nonumber
\\
&& L_4^{(2)} \equiv  \phi^{\mu} \phi_{\mu \nu}  \phi^{\nu \rho } \phi_{\rho}, \qquad L_5^{(2)} \equiv ( \phi^{\rho} \phi_{\rho \sigma} \phi^{\sigma})^2 \,,
\label{actions}
\eea
where $\phi_{\mu \nu}=\nabla_\nu \nabla_\mu \phi$, and $\phi_\mu = \nabla_\mu \phi$.  

\subsection{Degenerate and isokinetic theories}
The functions $F_2$ and $A_I$ satisfy three degeneracy conditions, as given in \cite{Langlois:2015cwa}, such that the DHOST action \eqref{dhost_action} propagates only one scalar degree of freedom, as well as two tensor modes.  Here we focus on the class of DHOST theories  that includes Horndeski and Beyond Horndeski theories \cite{Crisostomi:2016czh, Achour:2016rkg}, as it is the 
only phenomenologically viable one with real propagation speeds
(with no gradient instabilities at linear order) \cite{deRham:2016wji, Langlois:2017mxy}.
This class has been also shown to be stable under quantum corrections \cite{Santoni:2018rrx}.

Due to the three degeneracy conditions, the action is described in term of five independent functions: $F_0$, $F_1$, $F_2$ and two among the remaining five functions $A_I$.  We will take these to be $A_{1}$ and $A_3$. 
As can be seen directly from the 3+1 decomposition of the DHOST action given in \cite{Langlois:2015cwa}, the speed of propagation of gravitational waves, $c_g$, is equal to the speed of light $c=1$ only when $A_1=0$, in which case the weights of the kinetic term 
$K_{ij}K^{ij}$ and of the 3 dimensional scalar curvature ${}^{(3)}\! R$ are the same. The constraints 
inferred from the equality of the speed of light and that of  gravitational waves, following the observation of the neutron star merger GW170817,
have been discussed 
in \cite{Creminelli:2017sry,Ezquiaga:2017ekz,Sakstein:2017xjx,Baker:2017hug,Crisostomi:2017lbg,Langlois:2017dyl,Dima:2017pwp}. 
We will refer to the corresponding  subclass as ``isokinetic'' theories. However, for the purpose of generality, we do not impose this constraint for the moment.

\subsection{Cosmological action}

Now consider a homogeneous and isotropic universe with  a Friedmann-Lema\^itre-Robertson-Walker (FLRW) spatially flat metric
\bea
ds^2 \; = \; -N^2dt^2 + a^2 \delta_{ij} \, dx^i dx^j\, ,
\eea
where the lapse $N$ and the scale factor $a$ depend only on time, as does the scalar field $\phi$.  
Inserting this metric in the action (\ref{dhost_action}) and taking into account the degeneracy conditions,  (\ref{dhost_action})  
yields the homogeneous action
\bea
S_{\rm hom} \; &=& \; \int dt  \, N {a}^3 \left\{
-6 (F_2 - X A_1) \left[ \frac{\dot {a}}{N a} - {\cal V} \frac{\dot\phi}{N^2} \frac{d}{dt}\left(\frac{\dot\phi}{N}\right) \right]^2 \right.
\nonumber
\\
&& \qquad \qquad \qquad  \left. -3 (F_1+2F_{2\phi}) \frac{\dot {a}\, \dot\phi}{N^2a}  - F_1  \frac{1}{N} \frac{d}{dt}\left(\frac{\dot\phi}{N}\right)+ F_0 \right\}\, ,
\label{cosmodhost}
\eea
where $F_{2\phi} \equiv \partial_\phi F_2$, all the free functions now depend on $\phi(t)$ and ${X}\equiv -\dot\phi^2/N^2$, and ${\cal V}$ is given by
\bea
{\cal V} \; \equiv \; 
\frac{4 F_{2X}  +X A_3 -2A_1}{4(F_2 - X A_1)}\, .\label{Kev}
\eea
Note that, as a result of the degeneracy, the terms quadratic in $\dot a$ and $\ddot \phi$ combine into a square term in the first line of \eqref{cosmodhost}. This motivates the definition of a new scale factor, $b$, which absorbs the second time derivative of the scalar field,
\beq
\label{dot_b}
\frac{\dot b}{b}=\frac{\dot {a}}{a} - {\cal V} \frac{\dot\phi}{N} \frac{d}{dt}\left(\frac{\dot\phi}{N}\right)+\dots \, ,
\eeq
where the dots denote terms up to first order in time derivatives of $\phi$. Letting
\beq
a=e^{\lambda({X},\phi)} \, b \qquad \Rightarrow \qquad \frac{\dot a}{a}=\frac{\dot {b}}{b} +{\lambda_X}\dot{X}+ {\lambda_\phi}\dot\phi\,,
\label{HtoHb}
\eeq
it follows from ({\ref{dot_b})  that
\beq
\label{Lambda_X}
\lambda_X=-\frac12 {\cal V} = -\frac{4 F_{2X}  +X A_3 -2A_1}{8(F_2 - X A_1)}\, .
\eeq
Re-expressing the cosmological action (\ref{cosmodhost}) in terms of the new scale factor $b$ gives
\bea
S_{\rm hom} & = & \int dt \, N  b^3 \left[ \hat F_2(X,\phi) \frac{\dot{b}^2}{N^2b^2} + 
\hat F_1(X,\phi)\frac{\dot{b}\, \dot\phi}{N^2 b} + \hat F_0(X,\phi) +\hat G_1(X,\phi) \frac{1}{N} \frac{d}{dt}\left(\frac{\dot\phi}{N}\right) \right]\, ,
\label{actionHframe}
\eea
where it is straightforward to see that the functions $\hat F_I$ as well as $\hat G_1$ are given by
\bea
&&\hat F_2 \equiv -6e^{3\lambda} (F_2-XA_1)  \, , 
 \nonumber \\
&&
\hat F_1 \equiv -3  e^{3\lambda} \left[ F_1+2F_{2\phi} + 4 (F_2-XA_1) {\lambda_\phi}\right]\,, 
\\
&&
\hat G_1 \equiv -  e^{3\lambda} \left[ F_1+6X (F_1+2F_{2\phi}) {\lambda_X}\right]\, ,
 \nonumber \\
&&\hat F_0 \equiv  e^{3\lambda}\left[ F_0 +3X(F_1+2F_{2\phi}) {\lambda_\phi}  + 6 X (F_2-XA_1){\lambda_\phi^2}\right]  \,. \label{fff}
\eea

In general $\lambda$ is defined implicitly by the differential equation (\ref{Lambda_X}). 
In the special isokinetic case $A_1=0$, 
(\ref{Lambda_X}) reduces to
\beq
{\lambda_X}=-\frac{4 F_{2X}  +X A_3}{8F_2}.
\label{iso}
\eeq
In some cases, the integration can be done explicitly. In particular, when $X A_3$ is proportional to $F_{2X}$, namely $X A_3=-(4+8\mu) F_{2X}$
where $\mu$ is a constant, (\ref{iso}) 
gives $e^{\lambda}= \left(\frac{F_2}{M^2}\right)^\mu$,
where $M^2$ is an arbitrary (possibly $\phi$-dependent) function. We will consider this example in further detail below (see section \ref{section_application}).

From now on, we focus on shift symmetric theories in which all the functions appearing in the action depend only on $X$, and also $\lambda=\lambda(X)$. The reason is that, below, we will be interested in fixed points of the equations of motion. 
In non-shift symmetric theories, the fixed points are necessarily characterized by $\dot{\phi}=0=X$. In that case $\phi=$constant, and self-acceleration can only be generated by a potential for $\phi$. 
Here we are interested in situations in which dark energy is generated by the dynamics of the field itself.

\subsection{Equations of motion}
Starting from the action (\ref{actionHframe}), we now write down the equations of motion.  In the ``Horndeski frame'' with scale factor $b$, matter is not coupled minimally to the metric\footnote{If one describes matter by a $k$-essence like action $S_{matter} = \int d^4x \sqrt{-g} P(\sigma,\partial_\mu \sigma \partial^\mu \sigma)$ in the DHOST frame with scale factor $a$, then in the ``Horndeski'' frame, and in the case of the homogenous background, matter becomes conformally coupled to $b$, $S_{matter} = \int dt b^3 e^{3\lambda} P(\sigma, -\dot\sigma^2)$.}. 
How this non-minimal coupling appears in the  Friedmann equations is explained in the Appendix.

In terms of the original functions $F_I$ and $A_I$,
the Friedmann equations are given respectively by 
\bea
&&6\left[ F_2+(2 F_{2X} +6 \lambda_X F_2-3 A_1) X-2(A_{1X}+3\lambda_X A_1)X^2\right]H_b^2
\nonumber
\\
&&\quad +6 F_{1X} X H_b \, \dot\phi -2(F_{0X}+3 \lambda_X F_0) X +F_0=\left(1+6 w_m X\lambda_X\right) \rho_m \,,
\eea
and
\bea
&& 2 \left(F_2- X A_1\right)(2\dot H_b+3 H_b^2)+F_0 +2F_{1X} X \ddot\phi
\nonumber
\\
&&
\qquad +4\left[ F_{2X}+3 \lambda_X F_2- A_1- X( A_{1X} +3 \lambda_X A_1)\right]\dot X H_b =- P_m \, ,
\eea
where $\lambda_X$ is given in (\ref{Lambda_X}), and $\rho_m$ and $P_m$ are respectively the energy density and pressure 
defined in the physical frame. 
 In the following we assume an equation of state $P_m=w_m\, \rho_m$ with $w_m$ constant, which applies to the radiation dominated and matter dominated eras respectively.  Notice that this second equation contains $\ddot\phi$ and $\dot{H}_b$, and that $A_3$ only appears in the function $\lambda_X$.

 In the isokinetic case $A_1=0$, which we now consider for the remainder of this paper,  the above equations simplify
\bea
\label{Fried1}
{\cal E}_1&\equiv&6\left[ F_2+(2 F_{2X} +6 \lambda_X F_2) X\right]H_b^2+6 F_{1X} X H_b \, \dot\phi
\nonumber
\\
&&\quad  -2(F_{0X}+3 \lambda_X F_0) X +F_0=\left(1+6 w_m X\lambda_X\right) \rho_m \,,
\eea
and
\bea
\label{Fried2}
&&{\cal E}_2\equiv 2 F_2(2\dot H_b+3 H_b^2)+F_0 +2F_{1X} X \ddot\phi
 +4\left(F_{2X}+3 \lambda_X F_2\right)\dot X H_b =- P_m \, ,
\eea

In order to obtain the dynamical equation for $\phi$ -- that involves only $\ddot\phi$ but not $\dot H_b$ -- we first consider the combination 
\beq
w_m\, {\cal E}_1+(1+6w_m X \lambda_X){\cal E}_2=0,
\nonumber
\eeq
where the right hand side vanishes since $P_m=w_m \rho_m$.  This equation is linear in $\dot H_b$ and yields
\bea
\dot H_b&=&-\frac{1}{4 F_2(1+6w_m X \lambda_X)}\Big\{
(1+w_m)F_0 -2w_m XF_{0X} 
\nonumber
\\
&& \qquad +6H_b^2\left[2w_m X F_{2X}+(1+w_m+12w_m X\lambda_X)F_2\right] +2XF_{1X}(1+6w_m X \lambda_X)\ddot\phi
\nonumber
\\
&&
 \qquad 
+2H_b\left[3w_m XF_{1X}-4(F_{2X}+3 F_2\lambda_X)(1+6w_m X \lambda_X)\ddot\phi\right]\dot\phi
\Big\} \,.
\label{eq1}
\eea
We then substitute the above expression for $\dot H_b$ into the combination
\beq
(1+6w_m X \lambda_X)\left[\dot {\cal E}_1+3\left(H_b-2\lambda_X \dot\phi\ddot\phi\right)({\cal E}_1-(1+6w_m X \lambda_X){\cal E}_2)\right]-6 w_m  (X\lambda_X)\dot{}\, {\cal E}_1=0\,,
\label{eq2}
\eeq
which vanishes as a consequence of the conservation equation $\dot\rho_m+3H(\rho_m+P_m)=0$. This yields the equation of motion for the scalar field, which we do not write explicitly as it is quite involved.

\subsection{Self-accelerating attractors}
\label{section_dS}
The aim of this part is to determine the conditions under which it is possible to have 
self-accelerating cosmological expansion in DHOST theories. 

Self-accelerating de Sitter solutions can be easily identified by considering the equations of motion in the absence of matter, and imposing 
$\ddot\phi=0$ and $\dot H_b=0$.  Equation (\ref{Fried2}) directly yields
\beq
H_b=\sqrt{-\frac{F_0}{6F_2}} \equiv H_{dS}\,,
 \label{HbdS}
\eeq
which requires $F_0 F_2<0$. 
Substituting  in equation (\ref{Fried1}) gives
\beq
2\left[(F_0 F_2)_X +6\lambda_X F_0 F_2\right] ^2-3  F_0 F_2F_{1X}^2 X=0 \,,
\label{XdS}
\eeq
which is an equation for the constant $X \equiv X_{dS}$. 
We will study  in section \ref{section_application} the simple case in which $F_1=0$, and then $X_{dS}$ satisfies the equation
\beq
\label{eq_X_self_accelerating}
(F_0 F_2)_X +6\lambda_X F_0 F_2=0\,.
\eeq

In order to find the conditions under which the solution defined by (\ref{HbdS})  and \eqref{XdS} is an attractor,
we introduce the perturbations
\bea
x \; \equiv \; X - X_{dS} \, , \qquad
h \; \equiv \; H_b - H_{dS} \, .
\nonumber
\eea
Expanding (\ref{Fried1}) to  linear order gives $h$ in terms of $x$; and then substituting this expression into (\ref{Fried2}) gives a  first order equation for the perturbation $x(t)$,
\bea
{\cal A} \,  \dot x+  {\cal B} \,x \, = \, 0 \, ,
\nonumber
\eea
with ${\cal A}$ and ${\cal B}$ given by
\bea
\
{\cal A}  &\equiv &  -2 F_2^2 \left[2 X
   F_{0XX}+F_{0X} \left(6 X
   \lambda_X +1\right)+6 F_0 \left(2
   X \lambda_{XX} -6 X \lambda_X^2+\lambda_X
   \right)\right] \nonumber \\
   && +F_2 \left\{2
   F_0 \left[F_{2X} \left(18 X
   \lambda_X -1\right)-2 X
   F_{2XX}\right]+3 X^2
   F_{1X}^2\right\} +8 X
   F_0 F_{2X}^2 \nonumber \\
   && +2 \sqrt{6X
   F_0 F_2} \left\{F_2
   \left[X F_{1XX}+F_{1X}
   \left(1-6 X \lambda_X \right)\right]-2 X
   F_{1X} F_{2X}\right\} \,, \label{A} \\[2ex]
{\cal B}   & \equiv & -3 \sqrt{-X} \left\{F_{1X} \left[X
   F_2 F_{0X}+F_0
   \left(3 F_2-X
   F_{2X}\right)\right]+2 X F_0
   F_2 F_{1XX}\right\} \nonumber \\
   && -2\sqrt{-\frac{6F_0}{F_2}}
   \left\{F_2 \left[\left(X
   F_{0X}+F_0\right)
   F_{2X}+X F_0
   F_{2XX}\right] \nonumber \right. \\
   && \left. +F_2^2 \left[X
   F_{0XX}+F_{0X} \left(6 X
   \lambda_X +1\right)+6 F_0 \left(X
   \lambda_{XX} +\lambda_X \right)\right]-X
   F_0 F_{2X}^2\right\} \,, \label{B}
\eea
where $X$ is evaluated at the fixed point $X_{dS}$.  The condition for the self-accelerating universe to be an attractor is 
\beq
 {{\cal  A }} \, {{\cal B}} \,>0.
\label{stab-att}
\eeq
Notice that this condition is of course frame independent; had we worked from the start with $a$ and $H$ (rather than $b$ and $H_b$) the result would have been the same since, from (\ref{HtoHb}), at the fixed point $\dot{X}=0$ so that $H=H_b$.

We will consider this condition below (see section \ref{section_application}), in conjunction with 
the stability conditions coming from the study of linear cosmological perturbations to which we now turn.

\section{Stability of linear cosmological perturbations}
\label{section_perts}
In this section, we derive the second-order action for cosmological perturbations, and study it in the de Sitter phase (\ref{HbdS}), as well as in the case of a matter dominated universe. 
We begin by considering the case of an empty universe.

\subsection{Quadratic action in a cosmological background}

Contrary to the analysis of section \ref{section_dS}, here we work in the physical frame with scale factor~$a$. 
Linear cosmological perturbations around FLRW background have been studied for DHOST theories in \cite{deRham:2016wji, Langlois:2017mxy}. The quadratic action for the scalar perturbation $\zeta(t,x)$ (in the unitary gauge, and ignoring any matter contribution) was found to be \cite{Langlois:2017mxy}
\bea
S_{\rm quad}[\zeta] \; = \; \int d^3x \, dt \, a^3 \frac{M^2}{2} \left[ A_\zeta \, \dot \zeta^2 + B_\zeta \, \frac{(\partial_i \zeta)^2}{a^2}   \right] \, ,
\nonumber
\eea
where 
\bea
A_\zeta  &=&  \frac{1}{(1+\alpha_B -\dot \beta_1/H)^2} \left[\alpha_K + 6 \alpha_B^2 - \frac{6}{a^3 H^2 M^2} \frac{d}{dt}\left(a^3 H\, M^2 \alpha_B \beta_1\right)\right]  \,, \label{Azeta} \\ [2ex]
B_\zeta  &=& 2(1+\alpha_T) - \frac{2}{a M^2} \frac{d}{dt} \left[\frac{a M^2 \left(1+\alpha_H + \beta_1(1+\alpha_T) \right)}{H(1+\alpha_B) - \dot \beta_1}\right] \,, \label{Bzeta}
\eea
with $\alpha_T=0$ (since we focus here on the isokinetic case), and the parameters $\alpha_K$, $\alpha_B$, $\alpha_H$ and $\beta_1$ are given in terms of the free functions in (\ref{dhost_action}),
by
\bea
&& M^2 = 2 F_2 \,, \qquad \alpha_H= - \frac{2 X F_{2X}}{F_2} \, , \qquad
\beta_1 = \frac{X(4 F_{2X} + X A_3)}{4 F_2} \,, \\[2ex]
&& \alpha_B = -\frac{X \left(4 H F_{2X}+ 3 H X A_3 - 2 \sqrt{-X}
   F_{1X}\right)}{4 H
   F_2} \,, \\[2ex]
&& \alpha_K =  \frac{X}{2 H^2 F_2} \left\{2 \left[3 X^2 \left(\dot H+3 H^2\right)
   A_{3X}+2 X \left(F_{0XX}+6 \left(\dot H+2
   H^2\right) F_{2XX}\right) \nonumber \right. \right . \\
   && \left. \left. \qquad +F_{0X}+6 H
   \sqrt{-X} \left( X F_{1XX}- 
   F_{1X} \right) +6 \left(3 \dot H+2 H^2\right)
   F_{2X}\right] \nonumber \right. \\
   && \left. \qquad +3 X A_3 \left(5 \dot H+9
   H^2\right)\right\} \,.
\eea
The stability conditions (no ghost and no gradient instabilities) for linear cosmological perturbations are 
\bea
 A_\zeta  \, > \, 0 \, ,\qquad B_\zeta  \, < 0 \, .
  \label{vaccond}
\eea

At the de-Sitter fixed point  (\ref{HbdS}), the expressions for $A_\zeta$ and $B_\zeta$ are given by

\bea
A_\zeta  &=&  \Big\{18 X \left[2 F_2^2 \left(2 X
   F_{0XX}+F_{0X}+6 F_0 \left(2 X
   \lambda_{XX}-12 X \lambda_X^2+\lambda_X
   \right)\right) \right. \nonumber \\
   && \left. +F_2 \left(2 F_0
   F_{2X}+4 X F_0 \left(F_{2XX}-12
   F_{2X} \lambda_X \right)-3 X^2
   F_{1X}^2\right)-8 X F_0
   F_{2X}^2\right] \nonumber \\
   && +36 \sqrt{6} X \sqrt{X F_0
   F_2} \left[2 X F_{1X}
   F_{2X}+F_2 \left(F_{1X} \left(9 X
   \lambda_X -1\right)-X F_{1XX}\right)\right] \Big\} / (F_2 D^2) \,,
   \label{Azeta2} \\ [2ex]  
 B_\zeta  &=& \Big\{ X 6 \sqrt{6} X \sqrt{X F_0 F_2}
   F_{1X} \left[6 X F_{2X}+F_2
   \left(14 X \lambda_X +1\right)\right] \nonumber \\
   && -6 X
   \left[F_2 \left(80 X F_0 F_{2X}
   \lambda_X +8 F_0 F_{2X}+3 X^2
   F_{1X}^2\right) \right. \nonumber \\
   && \left. +16 X F_0
   F_{2X}^2+16 F_0 F_2^2 \lambda_X
    \left(6 X \lambda_X +1\right)\right] \Big\} / (F_2 D^2) \, ,
    \label{Bzeta2}
\eea
where $X$ is evaluated at the fixed point $X_{dS}$ given by the solution of (\ref{XdS}), and
\beq
D= X \left(2 \sqrt{6} \sqrt{-\frac{F_0}{F_2}}
   F_{2X}+3 \sqrt{-X} F_{1X}\right)+\sqrt{6}
   \sqrt{-F_0 F_2} \left(6 X \lambda_X
   +1\right) \,.
\eeq
We will illustrate these stability conditions, and their relation to (\ref{stab-att}), with a specific example in section \ref{section_application}.

\subsection{Including matter }

We now extend our analysis and consider the case in which matter is present. This is
interesting in order to see whether these theories could reproduce a history of the universe with a matter dominated era
before the self-accelerating one. We will show that this is indeed the case with the simple example of section \ref{section_application}.
 
We describe matter as a scalar field $\sigma(t,x)$, with a k-essence type action
\beq
S_m = \int d^4x \sqrt{-g} \, P(Y) \,, \qquad Y \equiv g^{\mu\nu} \partial_\mu \sigma \partial_\nu \sigma \,,
\nonumber
\eeq
which should be added to (\ref{dhost_action}).
To make contact with the perfect fluid description of matter, we identify the matter density $\rho_m$, the equation of state $P_m=w_m \rho_m$  and  the sound speed $c_m$
\beq
w_m = - \frac{P}{P+2 \dot \sigma_0^2 P_Y} \,, \qquad  c_m^2 = \frac{P_Y}{P_Y - 2 \dot \sigma_0^2 P_{YY}} \,,
\eeq
where $\sigma_0(t)$ denotes  the value of $\sigma$ on the cosmological background.

Following the same analysis as in \cite{Langlois:2017mxy} and after a long calculation, the quadratic action for the perturbations 
reduces to 
\beq
S_{\rm quad}[v] \; = \; \int d^3x \, dt \, a^3 \frac{M^2}{2} \left[ \dot v^\intercal \, {\bf K} \, \dot v + v^\intercal \, {\bf D} \, \dot v + \frac{1}{a^2} \, \partial_i v^\intercal \, {\bf L} \, \partial_i v +  v^\intercal \, {\bf M} \, v  \right] \, ,
\eeq
where $v^\intercal \equiv (\zeta, \delta \sigma)$ and
\bea  {\bf K} = \left(
\begin{matrix}
    A_\zeta + \frac{\rho_m(1+w_m)}{M^2 c_m^2 \left( H(1+\alpha_B) - \dot \beta_1\right)^2}    \qquad   &  \frac{ \rho_m (1+w_m) \left( 3 c_m^2 \beta_1 -1 \right)}{M^2 c_m^2 \left( H(1+\alpha_B) - \dot \beta_1\right) \dot \sigma_0} \\[4ex]
    \frac{ \rho_m (1+w_m) \left( 3 c_m^2 \beta_1 -1 \right)}{M^2 c_m^2 \left( H(1+\alpha_B) - \dot \beta_1\right) \dot \sigma_0}       & \frac{\rho_m(1+w_m)}{M^2 c_m^2 \dot \sigma_0^2}
\end{matrix} \right) \,, \qquad
{\bf D} = \left(
\begin{matrix}
    0   &  d \\[4ex]
    d       & 0
\end{matrix} \right) \,, \\[4ex]
{\bf L} = \left(
\begin{matrix}
    B_\zeta    &  \frac{ \rho_m (1+w_m) \left( 1 + \alpha_H + (1+ \alpha_T) \beta_1\right)}{M^2  \left( H(1+\alpha_B) - \dot \beta_1\right) \dot \sigma_0} \\[4ex]
    \frac{ \rho_m (1+w_m) \left( 1 + \alpha_H + (1+ \alpha_T) \beta_1\right)}{M^2  \left( H(1+\alpha_B) - \dot \beta_1\right) \dot \sigma_0}     & - \frac{\rho_m(1+w_m)}{M^2 \dot \sigma_0^2}
\end{matrix} \right) \,, \qquad 
{\bf M} = \left(
\begin{matrix}
    0   &  q \\[4ex]
    q       & m
\end{matrix} \right) \,. \\ \nb
\eea
Notice that the explicit forms of $m, q$ and $d$ (which are very involved) are not relevant for our purposes, and $A_\zeta$ and $B_\zeta$ are given in (\ref{Azeta}) and (\ref{Bzeta}) respectively.

The stability conditions, in order to avoid  ghost and gradient instabilities, translate, respectively, into the requirement that the eigenvalues of ${\bf K}$ should be positive and those of ${\bf L}$ should be negative. Let us study these conditions in a matter dominated era
where the radiation contribution to the matter content is supposed to be negligible.

For that purpose, we first denote these eigenvalues by
\beq
\text{Eigen}({\bf K}) = \left\{ \lambda_{{\bf K}_1} \,,\, \lambda_{{\bf K}_2} \right\} \,, \qquad
\text{Eigen}({\bf L}) = \left\{ \lambda_{{\bf L}_+} \,,\, \lambda_{{\bf L}_-} \right\} \,.
\nonumber
\eeq
In the matter dominated era, we assume $c_m \simeq w_m \ll 1$, so that the leading contribution in an expansion in $w_m$ gives
\bea
\lambda_{{\bf K}_1} &\simeq& \frac{A_\zeta M^2 \left( H(1+\alpha_B) - \dot \beta_1\right)^2 + 6 \rho_m \beta_1}{M^2 \left[ \left( H(1+\alpha_B) - \dot \beta_1\right)^2 + \dot \sigma_0^2 \right]} \,, \qquad  
\lambda_{{\bf K}_2} \simeq \frac{\rho_m}{M^2 w_m^2} \left[ \frac{1}{\left( H(1+\alpha_B) - \dot \beta_1\right)^2} + \frac{1}{\dot \sigma_0^2} \right]  \,, \nb \\ \label{eigenK} \\
\lambda_{{\bf L}_\pm} &\simeq& \frac{B_\zeta}{2} - \frac{1}{2M^2\dot \sigma_0^2} \left[ \rho_m \pm \sqrt{\frac{4\rho_m^2 \left( 1 + \alpha_H + (1+ \alpha_T) \beta_1\right)^2 \dot \sigma_0^2 }{\left( H(1+\alpha_B) - \dot \beta_1\right)^2} + \left( \rho_m + B_\zeta M^2 \dot \sigma_0^2 \right)^2 } \right] \,. \label{eigenL}
\eea

Concerning the eigenvalues of ${\bf K}$, one finds that $\lambda_{{\bf K}_2}$ in (\ref{eigenK}) is always positive whereas $\lambda_{{\bf K}_1}$ is positive only when its numerator is positive. 
The analysis of the eigenvalues of ${\bf L}$ is slightly more subtle because it involves the background evolution of $\sigma$.  

The background field equation of $\sigma$ is $\ddot \sigma_0 + 3 c_m^2 H \dot \sigma_0 = 0$ and therefore, in the matter dominated era where we assume that General Relativity is recovered (i.e. $H_m \simeq 2/3t$), this gives, 
assuming $c_m$ is constant,
\bea
\sigma_0(t) \propto \frac{t^{(1-2c_m^2)}}{1-2c_m^2} \, . 
\eea
At leading order (for small sound speed), it predicts a linear behavior in $t$ for $\sigma_0$. Using this property, as well as the dynamics of $H_m$ and $\rho_m$, in (\ref{eigenL}), we obtain the following expressions for the eigenvalues of ${\bf L}$ at the leading order in $H_m$ (i.e. small $t$)
\beq
\lambda_{{\bf L}_-} \simeq B_\zeta + \frac{3\left( 1 + \alpha_H + (1+ \alpha_T) \beta_1\right)^2}{2 M^2 (1+ \alpha_B)^2} \,, \qquad  
\lambda_{{\bf L}_+} \simeq - \frac{H_m^2}{M^2} - \frac{3\left( 1 + \alpha_H + (1+ \alpha_T) \beta_1\right)^2}{2 M^2 (1+ \alpha_B)^2} \,. \label{eigenL2}
\eeq
Hence, $\lambda_{{\bf L}_+}$ in the above expression is always negative, instead the requirement that $\lambda_{{\bf L}_-}$ is negative
leads to additional conditions. 
Applying to $\lambda_{{\bf K}_1}$ the same considerations that led from (\ref{eigenL}) to (\ref{eigenL2}), we obtain
\beq
\lambda_{{\bf K}_1} \simeq A_\zeta + \frac{9 \beta_1}{M^2 (1+\alpha_B)^2} \,.
\eeq

\medskip 

In conclusion, during the matter dominated phase of the universe, the conditions to avoid ghost and gradient instabilities read respectively  
\beq
A_\zeta + \frac{9 \beta_1}{M^2 (1+\alpha_B)^2} > 0  \qquad \text{and} \qquad B_\zeta + \frac{3\left( 1 + \alpha_H + (1+ \alpha_T) \beta_1\right)^2}{2 M^2 (1+ \alpha_B)^2} < 0 \,. \label{matterstab}
\eeq

\section{Application to a simple model}
\label{section_application}
Our aim in this section is to illustrate the different attractor and stability conditions derived so far. We propose a simple example in which these conditions turn out to be mutually compatible (which is not obvious, {\it a priori}).

We consider DHOST theories described by the functions
\beq
F_0= c_2 X\,, \qquad F_1=0\,,\qquad  F_2=\frac{M_0^2}{2}+c_4 X^2\,,\qquad A_3=-8c_4 - \beta,
\label{simple-example}
\eeq
parametrized by the  constant coefficients $M_0$, $c_2$, $c_4$ and $\beta$. The constant $M_0$ with a dimension of mass 
is clearly related to the Planck mass and the coefficients $c_4$ and $\beta$ incorporate the strong coupling scale $\Lambda_3$,
which is characteristic of this kind of theories \cite{Crisostomi:2017pjs}.

In principle the sign of $c_2$ is arbitrary and has not been fixed. However, as we will see below, in order for the different stability conditions (about the de Sitter solution) to hold simultaneously,  one must take $c_2>0$. In that case  $F_0<0$,  and we will assume that the second term in $F_2$ is always subdominant thus guaranteeing that $F_2>0$.

\subsection{Self-accelerating era}
The self-accelerating solutions are identified by solving (\ref{eq_X_self_accelerating}) which, in this case, leads to the fixed point
\beq
X_{dS}=-\sqrt{ - \frac{2}{3(4 c_4 + \beta)}}\, M_0 \,.
\label{solXds}
\eeq
The existence of a self-accelerating solution therefore imposes the condition $(4 c_4 + \beta) < 0$. 
The corresponding Hubble parameter  is given by (\ref{HbdS})
\beq
H_{dS}^2= - \sqrt{\frac23}\,\frac{c_2\sqrt{-(4c_4+\beta)}}{M_0(8c_4+3\,\beta)}\,,
\label{solHds}
\eeq
which is defined only if $(8 c_4+3\beta)<0$.
Using the expressions in equation (\ref{A}) and (\ref{B}), it is easy to verify that this solution is an attractor since
\beq
{\cal B}/{\cal A} = 3 H_{dS} > 0 \,.
\label{dsstability}
\eeq
Concerning the stability with respect to linear cosmological perturbations, we find that
\beq
\text{Sign}(A_\zeta)  =  \text{Sign}[ - c_2 (4 c_4 + \beta) ] \,, \qquad
\text{Sign}(B_\zeta)  =  \text{Sign}\left[ - \frac{c_2 c_4 (8 c_4 + \beta)}{ 8 c_4 + 3 \beta} \right]\,. \label{condvac}
\eeq
The condition $A_\zeta > 0$ is automatically satisfied, while $B_\zeta < 0$ implies $[c_4(8 c_4 + \beta)] < 0$, thus justifying the choice of $c_2>0$ (since we also have the condition $(8 c_4 + \beta) < 0$).

To summarize, at this stage, we have obtained the conditions  $c_2>0$, and  
\begin{itemize}
\item If $\beta>0$, then $c_4<0$ ;
\item  If $\beta<0$, then  $(4 c_4 + \beta) < 0$, and $c_4$ can be either positive or negative depending on the condition $[c_4(8 c_4 + \beta)]< 0$.
\end{itemize}

\subsection{Matter dominated era and scaling solutions}

In this section we focus on solutions which admit a matter dominated era before the de Sitter solution.   This situation has been analysed in \cite{DeFelice:2010pv} in the context of covariant Galileons theories, where it has been shown that `scaling solutions' play an important role, and also in \cite{Crisostomi:2017pjs} in an example of DHOST theories.   

To understand the properties of the scaling solutions, we first  rewrite the equations of motion (\ref{eq1}) and (\ref{eq2}) in the form\footnote{The analysis is simplified by working with the scale factor $b$ where all equations of motion are second order. Furthermore in these scaling solutions, $H=H_b$.}
 \begin{equation}\label{dynamicaleq}
\frac{d r_1}{d N} = f_1 (r_1, r_2)\,, \qquad 
\frac{d r_2}{d N} = f_2 (r_1, r_2)\,,
\end{equation}
by introducing new variables 
\begin{equation}
\label{newrvar}
r_1 =\frac{1}{\dot{\phi} H_b}\,, \qquad r_2= \dot{\phi}^4\,, 
\end{equation}
and using the e-folding $N= \ln b$ as a time variable. 
The Friedmann equation (\ref{Fried2}) gives a constraint equation $f_3(r_1, r_2)=0$. 
It is interesting to study the fixed points, from the point of view of the system \eqref{dynamicaleq}, which 
are determined by the conditions $f_1(r_1,r_2)=f_2(r_1, r_2)=0$.

First of all, a simple analysis shows that we recover, as expected, that the de Sitter solution is an attractor at late times. With these variables, the 
de Sitter spacetime is described by the fixed point solution
\begin{equation}\label{rdS}
r_{1 dS}=\sqrt{\frac{-3(8 c_4 + 3 \beta)}{2\,c_2}}\,, 
\qquad 
r_{2 dS} = - \frac{2 M_0^2}{3 ( 4 c_4 + \beta)} \, ,
\end{equation}
which obviously correspond to the self-accelerating solution given by equations (\ref{solXds}) and (\ref{solHds}). Linearising the equations of motion around this de Sitter fixed point, we find that the linear equations for $\delta r_1\equiv r_1 - r_{1 dS}$ and $\delta r_2\equiv r_2 - r_{2 dS}$ are given by
\begin{equation}
\frac{d \delta r_1}{d N} = - 3 \delta r_1, \quad 
\end{equation}
while the equation for $\delta r_2$ depends only on $\delta r_1$. 
Thus the de Sitter fixed point is stable in this theory, which is consistent with the de Sitter stability condition (\ref{dsstability}).   

Now, let us see whether there exists another fixed point (at early time) which would correspond  to a matter dominated era. In the case where matter has a constant equation of state 
$w_m = P_m/\rho_m$, we show that there is, indeed,  a new fixed point solution given by
\begin{equation}\label{rmatter}
r_{1m}= \sqrt{-\frac{3[16 c_4 + 3(1-w_m) \beta]}{ 4c_2}}\,, \qquad 
r_{2m}=0\,.
\end{equation}
By linearising the equations of motion \eqref{dynamicaleq} about this fixed point, we find that the linear equations 
for the perturbations $\delta r_1 \equiv r_1 - r_{1m}$ and $\delta r_2 \equiv r_2 - r_{2m}$ are
\begin{equation}
\frac{d \delta r_1}{d N} =  - \frac{3}{2} (3 + w_m) \delta r_1\,, \qquad 
\frac{d \delta r_2}{d N} = 6 (1+w_m) \delta r_2\,. 
\end{equation}
Hence, we see immediately that, for $w_m > -1$, this fixed point is a saddle point in which the branch $r_2$ is unstable. Therefore, 
given any initial conditions, the solution of the dynamical system approaches first the point \eqref{rmatter} at early times, and then 
it approaches the de Sitter fixed point \eqref{rdS} at late times. This is the scaling solution, and is illustrated in
Fig. \ref{Xi_evolution1} and Fig. \ref{Xi_evolution2} where we also show the transition between the saddle point and the de Sitter fixed point. 
\bfig
    \centering
    \includegraphics[width=	10cm]{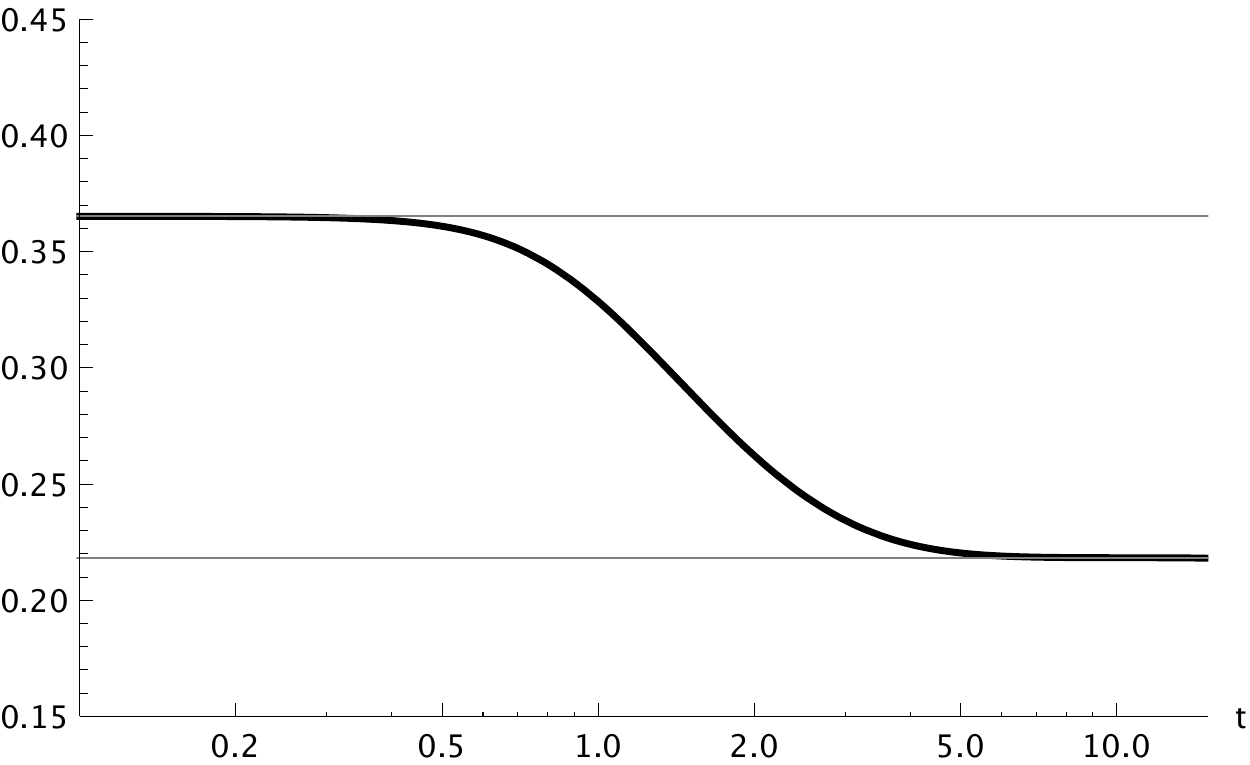} 
    \caption{$H \dot{\phi}$ as a function of time for the model with $c_2=1$, $c_4=1/2$ and  $\beta=-6$.  One can see the transition from the matter-dominated  scaling solution ($\xi_M=\sqrt{2/15}$) to the dS solution  ($\xi_{dS}=H_{\rm dS} \sqrt{-X_{\rm dS}}=1/\sqrt{21}$).}
    \label{Xi_evolution1}
\efig

\bfig
    \centering
    \includegraphics[width=	10cm]{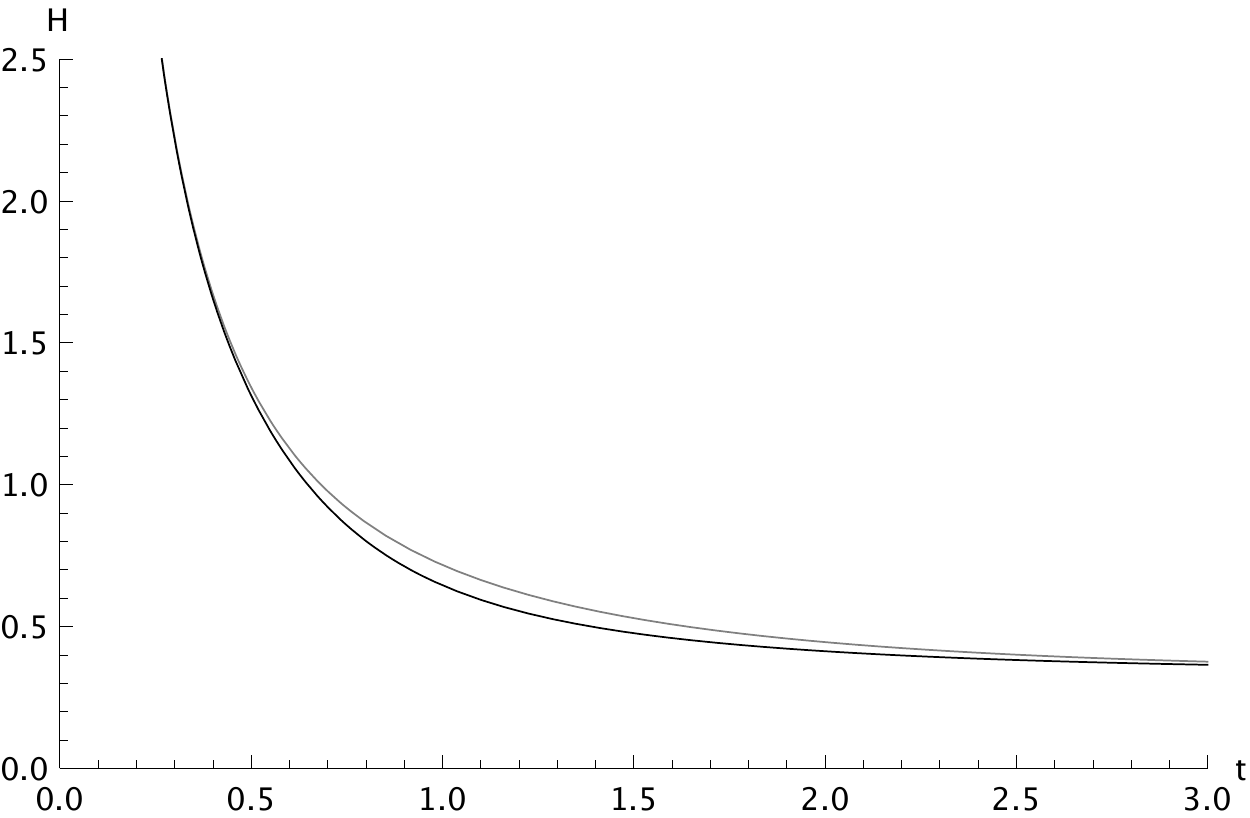} 
    \caption{The Hubble parameters $H$ (black) and $H_b$ (gray)  as functions of time for the model with $c_2=1$, $c_4=1/2$ and  $\beta=-6$.  The two Hubble parameters coincide in the matter dominated regime and in the de Sitter limit, but they differ during the transition between the two regimes.}
    \label{Xi_evolution2}
\efig

\medskip

It is interesting to propose an alternative way to recover the first equation of \eqref{rmatter}.  We will use this analysis in section 
\ref{Effective_EOS} to compute the effective equation of state of dark energy. Following  the definition of $r_1$ in \eqref{newrvar}, we set
\beq
\dot\phi=\frac{\xi}{H_b}\,,\qquad \Rightarrow \qquad \ddot\phi=\frac32(1+w_m)\xi,
\label{scalingsol}
\eeq
where $\xi$ is a constant that we want to determine. Notice that we have used  $H_b = {2}/{[3 (1+w_m)t]}$ in the previous equation, as it should be at early times (large $H_b$), hence $\dot\phi$ is small whereas $\ddot\phi$ is constant.  In order to find $\xi$, we substitute the equation (\ref{scalingsol}) into the  equation (\ref{eq2}) and consider that $H_b$ is large.  One then finds that the dominant term in the equation (which should vanish) is proportional to the combination
\beq
4c_2+3\left[16c_4+3(1-w_m)\beta\right]\xi^2\,.
\nb
\eeq
Hence, the scaling solutions have
\beq
\xi^2=- \frac{4c_2}{3\left[16c_4+3(1-w_m)\beta\right]}\,,
\label{xigeneral}
\eeq
thus confirming the result of \eqref{rmatter}.
In the matter era, this reduces to 
\beq
\label{xi_M}
\xi^2_M= - \frac{4c_2}{3\left(16c_4+3\beta\right)}\,,
\eeq
implying that $(16c_4+3\beta) < 0$.

\medskip

To finish, let us discuss the stability of the scaling solution with respect to cosmological perturbations, in the matter dominated era (early times).
For that purpose, we substitute this scaling solution in the conditions (\ref{matterstab}), under the same assumption that $H_b$ is large, and 
we find that the second condition is always satisfied since, at early times, the dominant term is $-3/2$. On the other hand, the first condition gives at the leading order the stability condition $(32c_4+3\beta) < 0$.

\medskip

Putting all together the conditions we have derived so far, we obtain the final constraint on the parameters of the model
\beq
\label{conds_model}
c_2 > 0 \,, \qquad c_4> 0 \,, \qquad  \beta < - \frac{32}{3} c_4 \,.
\eeq

\subsection{Effective equation of state for dark energy during matter domination}
\label{Effective_EOS}
It is instructive to calculate the effective equation of state for dark energy, which is subdominant in the matter dominated era. 
To do so, we rewrite the two Friedmann equations, in the physical frame, in the form
\beq
3 M_0^2 H^2=\rho_m+\rho_{\rm de}\,, \qquad M_0^2(2\dot H+3 H^2)=- P_m -P_{\rm de} \,,
 \eeq
where $\rho_{\rm de}$ and $P_{\rm de}$ are the effective energy density and pressure due to the modification of gravity.
In the matter dominated regime, we have $P_m=0$.  Using the two Friedmann equations (\ref{Fried1}) and (\ref{Fried2}), one finds 
  \beq
 \rho_{\rm de}= - \xi_M^2\left[c_2+ 30c_4 \xi_M^2\right]H^{-2} \,,
 \eeq
 and
 \beq
 P_{\rm de}=-\xi_M^2\left[c_2- 3\left( 8 c_4 - \frac{9}{4} \beta\right)\xi_M^2 \right]H^{-2}\,.
 \eeq
 Hence, the effective equation of state becomes
 \bea
 w_{\rm de}  =  \frac{c_2- 3\left( 8 c_4 - \frac{9}{4} \beta\right)\xi_M^2}{c_2+ 30c_4 \xi_M^2} \,,
 \eea
and substituting (\ref{xi_M}) yields 
\beq
\label{wde}
 w_{\rm de}=-2 \,.
 \eeq
This is expected from the scaling $\rho_{\rm de} \propto P_{\rm de} \propto H^{-2}$. If $\beta=0$, which corresponds to the case of Horndeski and Beyond Horndeski theories, one recovers the result obtained in \cite{DeFelice:2010pv}, which was shown to be problematic in explaining the observational data \cite{Nesseris:2010pc}. The models we have considered should however be seen as only illustrative because of their analytical simplicity, and one could extend them to overcome this difficulty. 
 
\section{Conclusion}
In this work, we have studied cosmological solutions and their stability properties in Degenerate Higher Order Scalar Tensor (DHOST) theories. Even though we have computed and presented the cosmological (background) equations for any quadratic DHOST theory, we have mainly studied cosmological properties of isokinetic DHOST theories for which gravity waves propagate at the light speed. Isokinetic theories are characterized by $A_1=A_2=0$ in the action \eqref{dhost_action}. 

We started with the study of  homogenous cosmological solutions. In this simplified framework, we see explicitly how degeneracy conditions enable us to recast higher order equations of motion (and then a higher order Lagrangian) into a second order system (associated to a Lagrangian with no higher derivatives) by a change of variables. This change of variables corresponds in fact to a change of frame by a disformal transformation of the metric in the fully covariant theory that transforms the DHOST action into a Horndeski action. 
We have derived the two Friedmann equations in this ``Horndeski frame''  (where matter is not coupled minimally to the metric) from which we have easily identified the conditions for having self-accelerating de Sitter solutions, and also the conditions for these solutions to be attractors.
Then, we have studied the stability of linear perturbations around these solutions in the language of the effective description of DHOST theories \cite{Langlois:2017mxy}.

In the last section, we have applied all these results to a simple class of theories, showing  that  de Sitter attractor conditions, no ghost and no gradient instabilities conditions -- both in the self-accelerating era and in the matter dominated era -- can be compatible. For the latter conditions, we have considered the case where matter is described in terms of a k-essence Lagrangian. In this paper we have extended the work initiated in \cite{DeFelice:2010pv} (for generalized galileons) and in \cite{Crisostomi:2017pjs} (for a simple class of DHOST theories) and found the existence of stable scaling solutions which interpolate between a matter dominated era ($w_m=0$) and a self-accelerating one. For these solutions, we computed the effective equation of state of dark energy in the matter dominated era. 

Besides the constraint on the anomalous speed of gravity from GW170817 (i.e. $A_1=A_2=0$), it was claimed very recently  that  the function $A_3$ should also vanish in order to prevent a catastrophic decay of gravitational waves into dark energy, an effect which would make GWs unobservable \cite{Creminelli:2018xsv}. This effect would further restrict the available class of DHOST models. However, a way out of all these constraints is to consider that the effective field theory described by DHOST is valid on cosmological scales but not necessarily on the much higher energy scales probed by LIGO/Virgo, as pointed out in \cite{deRham:2018red}. 
Moreover, a recent paper \cite{Copeland:2018yuh} suggested an interesting possibility to get around the above constraints by exploiting the dynamics of the scalar field itself, although this proposal ultimately fails when including the effect of large scale inhomogeneities.

\subsection*{Acknowledgments}
We thank Filippo Vernizzi for many enlightening discussions. We thank Shinji Tsujikawa for pointing out an error in the effective equation of state in the first version of the paper. MC is supported by the Labex P2IO and the Enhanced Eurotalents Fellowship. KK is supported by the UK STFC grant ST/N000668/1 and the European Research Council under the European Union's Horizon 2020 programme (grant agreement No.646702 ``CosTesGrav").

\appendix
\section{Non-minimal coupling of matter in the Horndeski frame}
Let us start with matter in the physical frame. As a direct consequence of the definition of the energy-momentum tensor
\beq
T^{\mu\nu}=\frac{2}{\sqrt{-g}}\frac{\delta S_m}{\delta g_{\mu\nu}}\,,
\eeq variation of the homogeneous matter action with respect to $N$ and to $a$  defines the matter energy density and pressure, respectively, according to
\beq
\label{matter_variation_physical}
\delta S_m =-a^3 \rho_m \delta N+ P_m \delta (a^3)\,.
\eeq

We now wish to derive the analog of (\ref{matter_variation_physical}) in the Horndeski frame. The simplest way to proceed is to use the explicit relation between the two scale factors $a$ and $b$, see (\ref{HtoHb}), to get
\beq
\frac{\delta a}{a}=-2X \lambda_X \frac{\delta N}{N}+\frac{\delta b}{b}\, .
\eeq
Substituting into (\ref{matter_variation_physical}), one finds
\beq
\delta S_m =b^3 \Lambda^3 \left[-\left(\rho_m +6X\lambda_X P_m\right)\delta N+ 3 P_m \frac{\delta b}{b}\right]\,.
\eeq
As a consequence, the energy density and pressure in the Horndeski frame are defined by
\beq
\tilde\rho_m=\left(1 +6X\lambda_X w_m\right)\Lambda^3\rho_m\,, \qquad \tilde P_m=\Lambda^3 P_m\,.
\eeq

\bibliographystyle{utphys}
\bibliography{cosmo_DHOST_biblio}

\end{document}